\documentclass[fleqn,usenatbib,useAMS]{mnras}

\usepackage{bm}
\usepackage{graphicx}	
\usepackage{amsmath}
\usepackage{amssymb}	
\usepackage{bm}
\usepackage{physics} 
\usepackage{multicol}
\usepackage{tabularx}
\usepackage{caption}
\usepackage{subcaption}
\usepackage{xcolor} 
\usepackage{soul}

\newcommand{\OmK}{\Omega_\text{K}}

\newcommand{\bv}{\mathbf{v}}

\DeclareUnicodeCharacter{2212}{-}

\title[]{On corrugation mode radial wavelengths of the vertical shear instability
}

\author[Dang, Cui \& Barraza-Alfaro]{
Yixuan Dang$^{1}$,
Can Cui$^{2,3}$\thanks{\href{mailto:can.cui@astro.utoronto.ca}{can.cui@astro.utoronto.ca}} and 
Marcelo Barraza-Alfaro$^{4}$
\\ \\
$^{1}$Department of Physics, University of Oxford, Parks Road, Oxford OX1 3PJ, UK \\
$^{2}$DAMTP, University of Cambridge, Wilberforce Road, Cambridge CB3 0WA, UK \\
$^{3}$Department of Astronomy and Astrophysics, University of Toronto, Toronto, ON M5S 3H4, Canada \\
$^{4}$Department of Earth, Atmospheric and Planetary Sciences, MIT, Cambridge, MA 02139, USA
}

\pubyear{2023}

\begin{document}
\label{firstpage}
\pagerange{\pageref{firstpage}--\pageref{lastpage}}
\maketitle

\begin{abstract}

The vertical shear instability (VSI) is a promising mechanism to drive turbulence in protoplanetary disks. Numerical simulations in the literature demonstrate that the VSI non-linear saturation is predominated by the linear corrugation modes. These modes possess vertical wavelengths crucially longer than radial wavelengths. 
This paper aims to investigate the natural radial wavelength of corrugation modes upon VSI saturation, by a series of numerical simulations conducted in Athena++ at different grid resolutions, disk aspect ratios, and viscosity parameterized by $\nu$. We find a sign of convergence emerges at 64 cells per gas scale height for fiducial simulations, below which a continuous reduction of wavelengths with grid resolution is observed. Synthetic ALMA molecular line observations of $^{12}\rm CO(2-1)$ are performed to inspect the observability of the corrugation modes feature, which is significantly diminished with more than 32 cells per scale height. Flared and viscous disks, exhibiting longer saturation wavelengths, may mitigate the observational difficulty.

\end{abstract}


\section{Introduction}\label{in}

The vertical shear instability (VSI) is a promising hydrodynamic mechanism to drive turbulence in protoplanetary disks (PPDs). It is the disk analogue of the Goldreich--Schubert--Fricke instability \citep{gs67,fricke68}. The instability criteria involve a vertical gradient of angular velocity due to baroclinity, and fast cooling to overcome the vertical stabilizing buoyancy. Early studies have illustrated the linear behavior \citep[e.g.,][]{gs67,ly15,lp18} and non-linear evolution \citep[e.g.,][]{nelson_etal13,sk14} of the VSI. The instability properties in weakly ionized protoplanetary disks have later been investigated \citep[e.g.,][]{cl21,lk22,cb20,cb21,cb22}. Recent radiation-hydrodynamic VSI simulations that incorporate two-moment methods have also been performed \citep{fuksman+23a,fuksman+23b}.

The non-linear saturation of the VSI exhibits remarkably coherent motions. This can be understood as radially traveling  and vertically standing waves \citep{cl22}, likely inherited from the linear corrugation modes \citep{nelson_etal13,bl15}. 
The characteristic feature of such modes possesses vertical wavelengths significantly longer than radial wavelengths. 
Local shearing sheet linear analysis suggests that the fastest growing modes have radial wavelength over vertical wavelength on the order of disk aspect ratio $h$ \citep{lp18}.
Numerical simulations demonstrate that the vertical wavelength of corrugation modes is a few gas scale height, whereas the measurement of radial wavelengths is less focused. 
\citet{sk14} has been the first to conduct resolution study on the radial wavelength of the corrugation modes. They found wavelength continuously reduces with grid resolution. With the highest resolution of about $60$ cells per scale height, the saturation of radial wavelength was not observed.

The coherent saturation feature renders VSI a potential candidate to be readily observed. \citet{flock_etal17} computed the synthetic images of dust continuum at a wavelength of 0.87 mm from the radiation hydrodynamic VSI simulations. The grid resolution adopted there is about 70 cells per scale height. Convolved with a 2D Gaussian filter to mimic ALMA observations, they found that the VSI feature induced by corrugation modes are almost smoothed out. \citet{marcelo+21} presented the synthetic CO rotational emission lines based on gas velocity structures induced by the corrugation modes. \citet{marcelo+23} further investigated the CO kinematic in the VSI turbulent disks embedded by a massive planet. A relatively low resolution of less than 20 cells per scale height was employed. It is likely that, as a result of this low resolution, the corrugation feature is not washed out in \citet{marcelo+21,marcelo+23}, as opposed to \citet{flock_etal17}.  

Besides the potential observational impact, corrugation mode wavelengths are crucial for the inertial-wave interactions that is proposed as a final state of the VSI. \citet{cl22} indicated that the coherent saturation feature may be unstable to a parametric instability induced by three-inertial-wave resonant interactions. This parametric instability can initiate an inertial-wave turbulent cascade, and hence transfer the energy from the large-scale corrugation modes to the small scales. Linear theory predicted that the length scale of the inertial waves excited by the parametric instability depends on the radial scale of the corrugation modes. Owing to the high resolution required to resolve the parametric instability, it has only demonstrated by local simulations with spectral code. Quantifying the corrugation mode wavelengths can clarify the requested resolution for observing the parametric instability in future global numerical simulations. 

In this paper, we investigate the radial wavelength of the corrugation modes by conducting 2D global numerical simulations in Athena++, with different disk aspect ratios and viscosity paramerized by $\nu$. The questions we seek to answer are: can radial wavelength of the corrugation modes saturate with grid resolution; can saturated wavelength be resolved given current observational capabilities. Our results yield a positive answer to the first question with good convergence found at a radial resolution of 64 cells per scale height for the fiducial simulations. Furthermore, the synthetic ALMA observations show that, at such radial resolution, the wavelength of the VSI pattern is almost diminished. These results flag the potential difficulty in recognizing VSI patterns in realistic molecular line observations, but can be mitigated by higher disk aspect ratio or viscosity.

The paper is organized as follows. In \S\ref{sec:method}, we present the dynamical equations, disk model, and a list of parameters employed in the numerical simulations. In \S\ref{sec:results}, we measure the corrugation mode radial wavelengths from the simulations and present the synthetic $^{12}$CO(2−1) line observations. We summarize and discuss the main findings in \S\ref{sec:cd}.

\section{Method}\label{sec:method}

\subsection{Dynamical equations}\label{sec:equations}

We use the grid-based high-order Godunov MHD code Athena++ to carry out numerical simulations in this work \citep{stone+20}. The mass, momentum, and energy equation in the conservative form read
\begin{equation}
\pdv{\rho}{t} + \nabla\cdot{(\rho \vb{v})} = 0, 
\end{equation}
\begin{equation}
\pdv{(\rho\vb{v})}{t}+\nabla\cdot(\rho\vb{v}\vb{v}-P\vb{I} - \btau) = - \rho\nabla{\Phi},  
\end{equation}
\begin{equation}
\pdv{E}{t}+\nabla\cdot{(E+P)\vb{v}} =-\rho(\vb{v}\cdot \nabla \Phi) -\Lambda_\mathrm{c},
\label{eq:energy}
\end{equation}
and the viscous stress tensor $\btau$ is
\begin{equation}
\btau=\eta[\nabla\bv+(\nabla\bv)^\mathrm{T}]-\frac{2}{3}\eta(\nabla\cdot\bv)\vb{I}.
\end{equation} 
Here $\vb{v}$, $\rho$, and $P$ are gas velocity, density, and pressure, respectively. The identity tensor is denoted by $\vb{I}$. The dynamical viscosity is denoted by $\eta=\rho\nu$, and $\nu$ is the kinematic viscosity. The total energy density is $E = \epsilon+\rho v^2/2$, where $\epsilon$ is the internal energy density and is related to the gas pressure by an ideal gas equation of state $P=({\gamma-1})\epsilon$. We adopt adiabatic index $\gamma=7/5$ for molecular gas. The cooling term $\Lambda_{\rm c}$ in the last equality will be elaborated in \S\ref{sec:model}. The gravitational potential of the protostar is implemented as a source term and given by $\Phi=-GM_\star/r$, with stellar mass $M_\star$. The simulations are conducted in spherical polar coordinates $(r, \theta, \phi)$, and cylindrical coordinates $(R, z, \phi)$ are used to improve presentation.

\subsection{Disk model and thermodynamic evolution}\label{sec:model}

We employ radial power-law temperature and density profiles as the initial condition \citep[e.g.,][]{nelson_etal13},
\begin{equation}\label{temperature profile}
    T(R)=T_0 \left( \frac{R}{R_0} \right)^{q_T},
\end{equation}
\begin{equation}\label{density profile}
    \rho(R,z=0)=\rho_0 \left( \frac{R}{R_0} \right)^{q_D}.
\end{equation}
In eq. (\ref{temperature profile}), we assume that temperature is constant on the cylinder (vertically isothermal). Parameters $q_T$ and $q_D$ describe the steepness of the power-law profiles, which the values can be found in Table \ref{table:1}.
To fully specify the density distribution of the disk, we solve the momentum equation in $R$ and $z$,
\begin{equation}\label{force equil 1}
    R\Omega^2-\frac{GMR}{r^3}-\frac{1}{\rho}\frac{\partial P}{\partial R}=0,
\end{equation}
\begin{equation}\label{force equil 2}
    -\frac{GMz}{r^3}-\frac{1}{\rho}\frac{\partial P}{\partial z}=0.
\end{equation}
Eqs. (\ref{force equil 1}) and (\ref{force equil 2}) give the density distribution,
\begin{equation}\label{density solution}
    \rho(R,z)=\rho_0 \left( \frac{R}{R_0} \right)^{q_D} \exp[\frac{GM}{c_s^2}\bigg(\frac{1}{r}-\frac{1}{R}\bigg)].
\end{equation}
Meanwhile, we can solve for the angular velocity profile,
\begin{equation}\label{angvel solution}
    \Omega(R,z)=\OmK\left[(q_T+q_D)h^2+1+q_T-\frac{q_TR}{r}\right]^{\frac{1}{2}},
\end{equation}
where the Keplerian angular velocity is defined as $\OmK=\sqrt{GM/R^3}$, and $H$ is the gas scale height. Eqs. (\ref{temperature profile}), (\ref{density profile}), (\ref{density solution}) and (\ref{angvel solution}) fully describe the initial disk model.

We relax the temperature $T(t)$ to its initial equilibrium value $T_0$ at each location in the disk by a relaxation timescale $\tau$,
\begin{equation}\label{thermal relaxation}
    \frac{\mathrm{d}T}{\mathrm{d}t}=-\frac{T-T_0}{\tau},
\end{equation}
where $\tau$ is a fraction of the local Keplerian orbital period $P=2\pi/\OmK$. We take $\tau=10^{-20}P$ in the simulations (locally isothermal), and adjust the amount of temperature after each simulation time step $\Delta t$ by 
\begin{equation}
\Delta T=(T_0-T)[1-\exp(-\frac{\Delta t}{\tau})].
\end{equation}

\begin{table}
\centering
\begin{tabular}{c c} 
 \hline
 parameters & values  \\ [0.5ex] 
 \hline
 $R_{in}/R_{0}, R_{out}/R_{0}$ & $1.0, 10.0 $ \\ 
 $\theta_{in,out}$             & $1.32-1.82$ \\ 
 $GM$       & 1.0  \\ 
 $R_{0}$    & 1.0  \\ 
 $\rho_{0}$ & 1.0 \\ 
 $q_T$      & -1.0  \\
 $q_D$      & -1.5  \\ 
 $\tau/P$   & $10^{-20}$  \\ 
 noise      & 1\%  \\[1ex] 
 \hline
\end{tabular}
\caption{List of parameters employed in simulations. From top to bottom: radial domain, meridional domain, gravitational units, reference radius, reference density, density power-law index, temperature power-law index, thermal relaxation timescale, and amplitude of velocity noise in units of local sound speed.}
\label{table:1}
\end{table}

\begin{table*}
\centering
\begin{tabular}{l|@{\hskip 0.2in} l l l l l c} 
 \hline
 model &  grid size ($N_{r}\times N_{\theta}\times N_{\phi}$) & cells per $H$ & grid ratio in $R$  & $h$ &$\nu$ & run time ($P_0$) \\ [0.5ex] 
 \hline
 cph8& $380  \times80\times1$ & $8\times8$ & 1.006 & 0.05 & 0 & $1.5\times10^3$ \\ 
 cph16& $750 \times160\times1$ & $16\times16$ & 1.003 & 0.05 & 0 & $1.5\times10^3$\\ 
 cph32 & $1500\times320\times1$ & $32\times32$ & 1.0015  & 0.05 & 0 & $1.5\times10^3$\\ 
 cph64& $3000\times640\times1$ & $64\times64$ & 1.00077  & 0.05  & 0 & $1.5\times10^3$\\  
 cph98& $4620\times980\times1$ & $98\times98$ & 1.0005   & 0.05 & 0 & $1.5\times10^3$\\
 cph128 & $6020\times1280\times1$ & $128\times128$ & 1.00038 & 0.05 & 0 & $1.5\times10^3$ \\
 \hline
 cph\texttt{\textunderscore}hr0.1 & -- & --& -- & 0.1 & 0 & $10^3$\\
 cph\texttt{\textunderscore}visc & -- & --& -- & 0.05 & $10^{-6}$  & $1.5\times10^3$\\

\hline
\end{tabular}
\caption{List of parameters including grid resolution, disk aspect ratio ($h$), kinematic viscosity ($\nu$) and run time for each simulation model.}
\label{table:2}
\end{table*}

\subsection{Simulation setup}\label{setup}

We conduct three groups of simulations with different disk aspect ratios and viscosity, denoted by \texttt{cph}, \texttt{cph\_hr0.1} and \texttt{cph\_visc}. Within each group, simulations share the same physical parameters, but are carried out under different grid resolutions. Table \ref{table:1} presents the values of physical parameters used, and Table \ref{table:2} shows the resolution setup of each model. The models are named after their cells-per-scale-height values. 

Simulations in the fiducial models \texttt{cph} are carried out under $\nu=0$ and $h=0.05$, while models \texttt{cph\_hr0.1} are under the conditions of $\nu=0$ and an increased $h=0.1$, and models \texttt{cph\_visc} has constant viscosity of $\nu=10^{-6}$ and $h=0.05$. A constant kinematic viscosity gives $\alpha\propto R^{-1/2}$. Hence, the equivalent $\alpha$ values are $4\times10^{-4},~1.7\times10^{-4},~1.2\times10^{-4}$ for $h=0.05$, or, $10^{-4},~4.4\times10^{-5},~3\times10^{-5}$ for $h=0.1$ at $R=1,~5,~10$.
We vary the aspect ratio because $h=0.05$ might be small for a location beyond $100~$au of flared protoplanetary disks \citep[e.g.,][]{zhang2021}, and because of the potential correlation between the radial wavelength and aspect ratio revealed by linear theory \citep{lp18}. Models \texttt{cph\_visc} are introduced as we are aware of the potential of other viscous processes in the disk that may coexist with the VSI.

\section{Simulation Results and Synthetic Line Observations} \label{sec:results}

\begin{figure*}
    \includegraphics[width=1\textwidth]{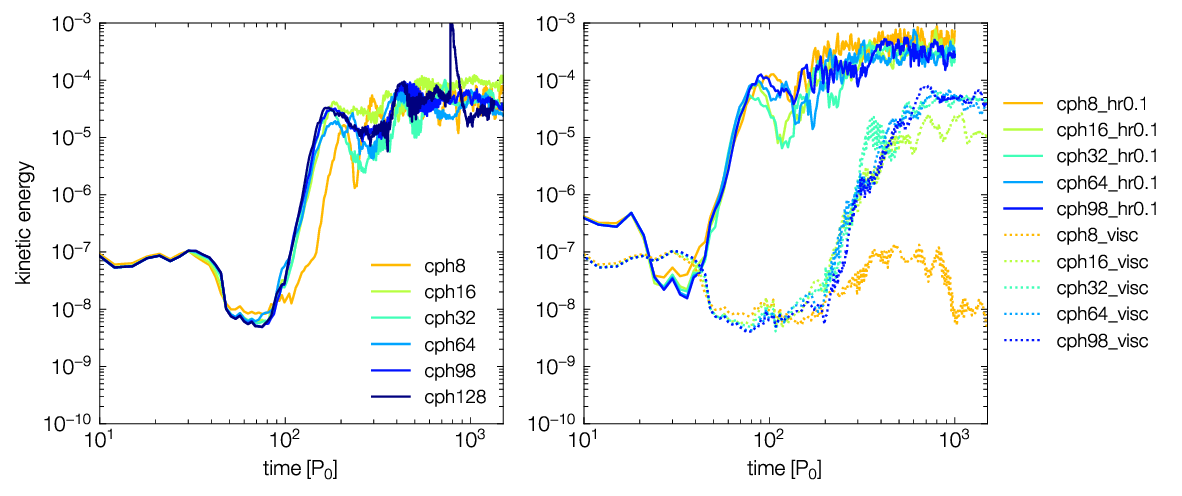}
    \caption{Sum of radial and meridional kinetic energy normalized by initial azimuthal kinetic energy as a function of time in units of $P_0$.}
    \label{fig:ke}    
\end{figure*}

\begin{figure*}
    \includegraphics[width=0.97\textwidth]{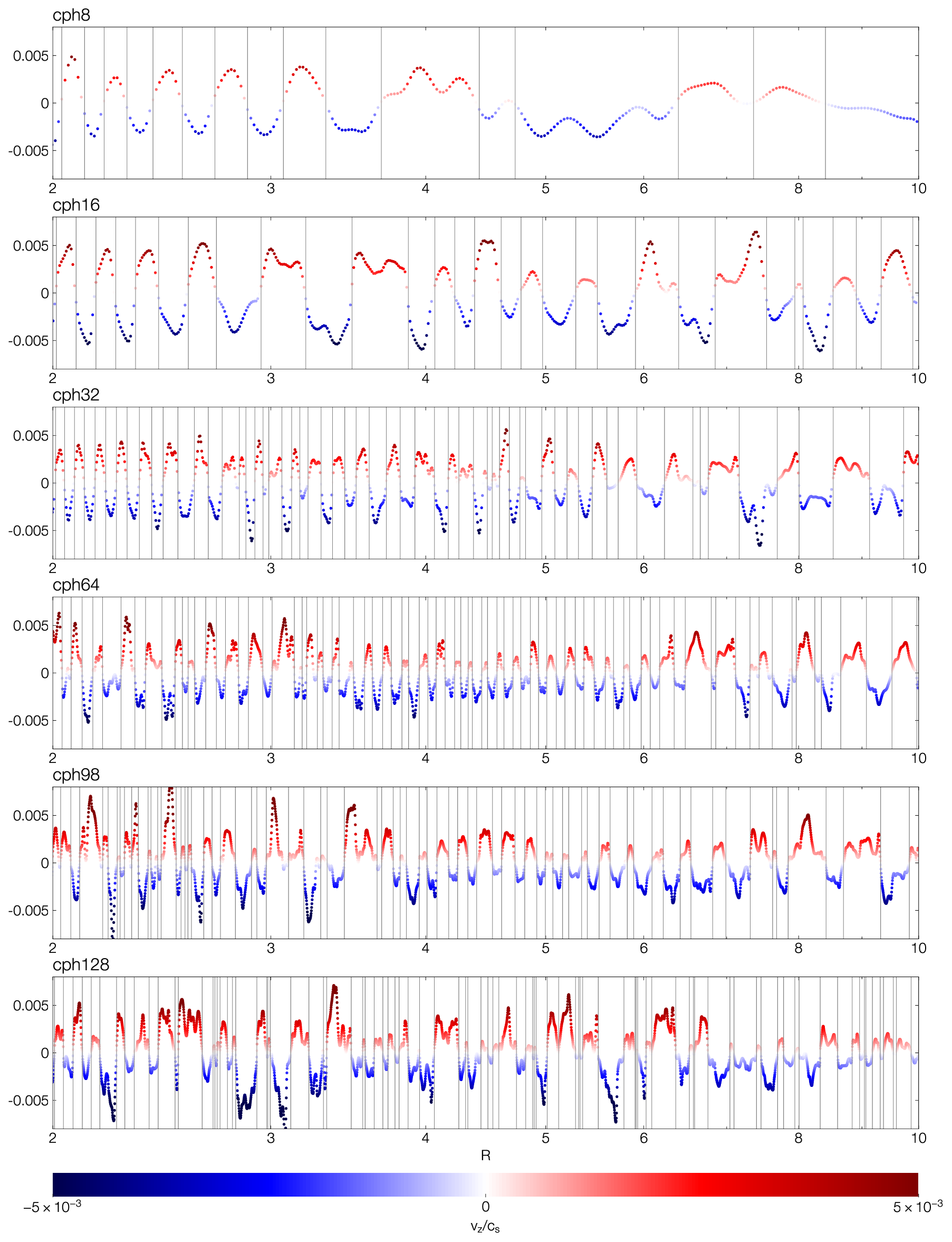}
    \caption{Midplane vertical velocity divided by local sound speed $v_z/c_s$ at $t=1300P_0$ for fiducial models. From top to bottom: cph8, cph16, cph32, cph64, cph98, cph128. The vertical gray lines indicate the radial locations of $v_z$ sign change.}
    \label{fig:vz}    
\end{figure*}

\begin{figure*}
    \includegraphics[width=0.95\textwidth]{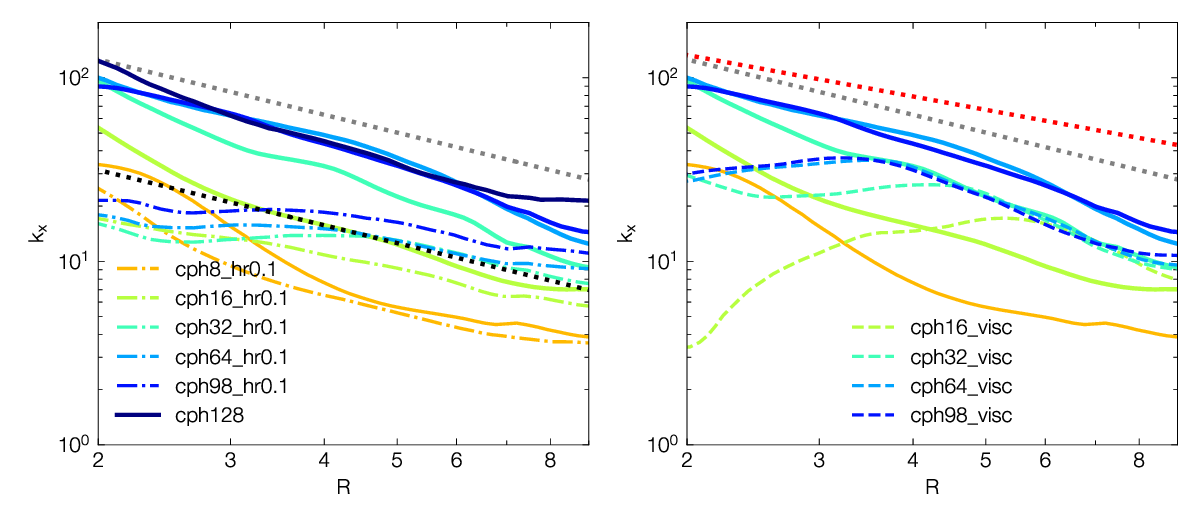}
    \caption{
    Wavenumber $k_x$ as a function of $R$ averaged over $1000-1500P_0$ for \texttt{cph} (solid) and \texttt{cph\_visc} (dashed), and over $500-1000P_0$ for \texttt{cph\_hr0.1} (dash-dotted). The curves are smoothed for better presentation. Dotted grey ($h=0.05$) and black ($h=0.1$) lines denote the predictions by linear theory \citep{lp18}. Dotted red lines denote the maximum radial wavenumber set by viscosity from the prediction of \citet{ly15}.
    }
    \label{fig:wn}    
\end{figure*}

\begin{figure*}
    \centering
    \includegraphics[width=0.97\textwidth]{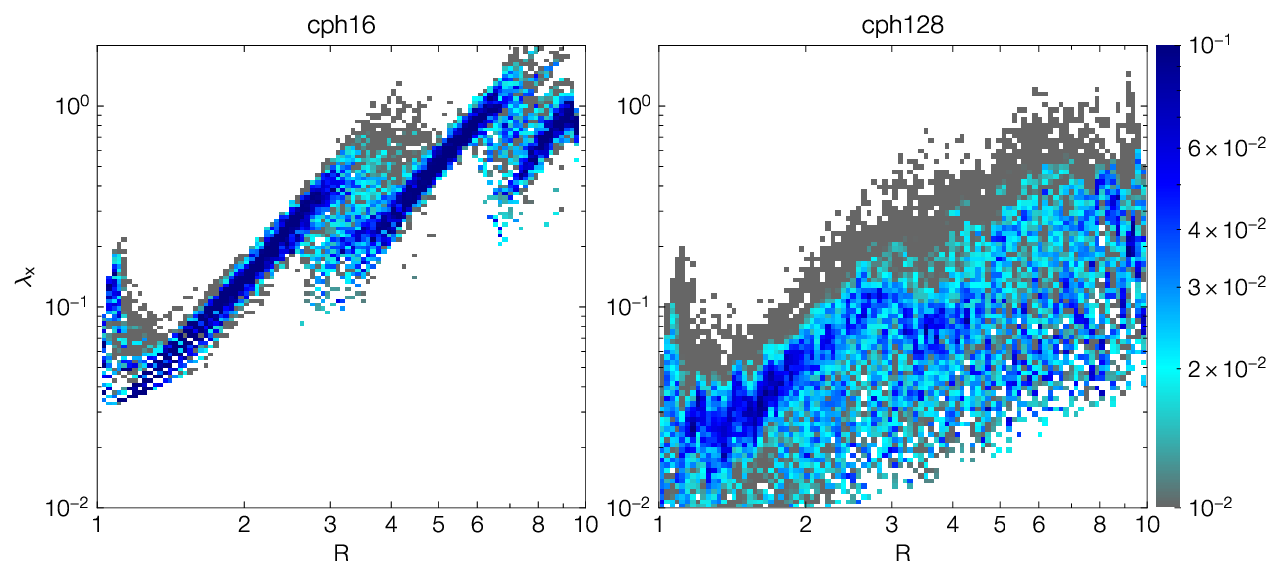}
    \caption{Wavelengths of corrugation modes versus radius over a time interval from $1000$ to $1500P_0$ of \texttt{cph16} and \texttt{cph128}. Colours in logarithmic scale are the probability that a specific wavelength can occur at a fixed radial position bin. 
    }
    \label{fig:8}    
\end{figure*}

In this section, we measure the radial wavelength of corrugation modes from numerical simulations (\S\ref{sec:3.1}), and present the $^{12}$CO(2−1) synthetic line observations to inspect the effect of grid resolution (\S\ref{sec:SLO}).

\subsection{Simulation results} \label{sec:3.1}

Figure \ref{fig:ke} computes the volume-integrated radial and meridional kinetic energies, normalized by the azimuthal kinetic energy in the initial state of the Keplerian motion,
\begin{equation}
\mathrm{E_k}=\frac{ \frac{1}{2}\int \rho v_{r}^{2}\ dV + \frac{1}{2}\int\rho v_{\theta}^{2}\ dV}{\frac{1}{2}\int \rho v_{\phi}^{2}\ dV}.
\label{KE}
\end{equation}
The kinetic energy is measured in a domain of $r\in[6,8]$ and $\theta=\pi/2\pm0.2$.
For fiducial simulations (\texttt{cph}; left panel), we note that after $\sim500P_0$, where $P_0=P(R_0)$, the kinetic energies for all models stay at a constant with small fluctuations. We take this as the sign of saturation and use time intervals from $1000$ to $1500P_0$ to conduct analysis, for which the radial wavelengths are considered fully saturated. 
We also study the evolution of kinetic energy for models \texttt{cph\_hr0.1} and \texttt{cph\_visc} in comparison. The time interval chosen to conduct analyses for \texttt{cph\_hr0.1} is from $500$ to $1000P_0$. Models \texttt{cph\_visc} saturate slower due to the viscous damping, and a time interval of $1000$ to $1500P_0$ is chosen.

Figure \ref{fig:vz} shows vertical velocities at the midplane divided by the local sound speed at $t=1300P_0$ of fiducial models \texttt{cph}. 
To measure the radial wavelenth, we first locate the radii where vertical velocity at midplane changes sign. These are marked as grey vertical lines in Figure \ref{fig:vz}. We record the distance between two consecutive lines as the start and the end of half of a wavelength. Multiplying this distance by a factor of two yields the wavelength at each radial location. From Figure \ref{fig:vz}, we can already observe a clear trend of decreasing wavelength as resolution increases. 

Figure \ref{fig:wn} plots the radial profile of wavenumber $k_x=2\pi/\lambda_x$, where $\lambda_x$ is the radial wavelength. Fiducial models \texttt{cph} are denoted by solid lines, models \texttt{cph\_hr0.1} by dash-dotted lines, and models \texttt{cph\_visc} by dashes lines. 
On the left panel, solid lines directly compare the radial wavelengths at different resolutions in models \texttt{cph}, confirming the observation in Figure \ref{fig:vz}. At a resolution of 64 cells per $H$ or above, there exists a clear trace of convergence to a minimum wavelength, in close proximity to the predication by linear theory $\lambda_x/\lambda_z\sim h$ \citep{lp18}, shown as grey dotted line, and we take $\lambda_z\sim 10H$ corresponding to the vertical domain of the simulation. Opposed to our findings, \citet{sk14} did not observe a converging wavelength of corrugation modes, which could be attributed to their maximum resolution of $\sim 60$ cells per $H$ adopted.
The same convergence pattern is spotted for models \texttt{cph\_hr0.1}, though an overall longer wavelengths are obtained for higher disk aspect ratio. This is consistent with the linear theory shown as black dotted line. 

On the right panel of Figure \ref{fig:wn}, models \texttt{cph\_visc} show that viscosity erases unstable modes of short wavelengths, giving lower limits on $\lambda_x$. Models \texttt{cph\_visc} have longer saturated wavelengths overall compared to models \texttt{cph} due to the viscous damping.
The maximum wavenumber set by viscosity can be estimated by $k_x^2\le |q_T|h\Omega/\nu$ \citep{ly15}. We plot the theoretically predicted maximum $k_x$ as a red dotted line in the right panel of Figure \ref{fig:wn}. It is clear that all the wavenumbers obtained for models \texttt{cph\_visc} are well below it.
Note that the wavenumber for \texttt{cph8\_visc} cannot be reliably calculated because VSI is almost erased by viscosity at this low resolution.
\citet{sk14} also obtained a convergence of wavelengths when employing a lower kinematic viscosity $\nu=5\times10^{-7}$. 

Figure \ref{fig:8} shows corrugation modes wavelengths in a time interval from $1000$ to $1500P_0$. Here, we present \texttt{cph16} and \texttt{cph128} as examples. We bin the wavelengths and radial positions. The data points are collected and accumulated over each snapshot in the selected time interval. The color in logarithmic scale denotes the probability for a wavelength to occur at a fixed radial bin. At a given radial position bin, it is calculated by taking the ratio of data points collected in each wavelength bin to the total data points collected at this radial bin over all wavelength bins. 

Figure \ref{fig:8} allows us to have an overview of the spatial distribution of wavelength for individual snapshots such that the trend obtained from averaged data in Figure \ref{fig:wn} can be confirmed to represent the situation at all snapshots. 
For model \texttt{cph16}, we observe that the wavelength of corrugation modes does not have a significant time dependence once saturated. The distribution of wavelength is quite concentrated. We also see more clearly how wavelengths increase with radius. The three wave zones appeared can be explained by linear theory developed in \citet{svanberg+22}.
For model \texttt{cph128}, the wavelengths are notably more scattered at a fixed radius, but they generally show a shift towards shorter wavelengths compared to model \texttt{cph16}. We note that this scattered wavelengths pattern is more significant with higher grid resolution.

\subsection{Synthetic line observations}\label{sec:SLO}

In order to inspect how the radial wavelength modifies the observables, we generate the $^{12}$CO(2−1) synthesized line observations using the fiducial simulation data (\texttt{cph}). 
We post-processed the simulation outputs with the radiative transfer code \textsc{RADMC-3D} \citep{Dullemond2012} version 2.0. For this procedure, we follow an analogous approach as presented in \cite{marcelo+23} (see also \cite{marcelo+21}, and references therein).
We interpolate all simulations to a common grid resolution of $N_r \times N_{\theta}=512 \times 128$ and include the azimuthal direction by assuming axisymmetry, using $N_{\phi}=512$.
We assume a disk perfectly face-on, following the parameters of TW Hydrae protoplanetary disk, that is, a central star of $0.88 M_{\odot}$ \citep{Andrews2012, Huang2018}, and a distance to the object of 60.1 pc \citep{Gaia2018}. We scale the simulations such that the disk extends radially from 25 to 250 au, so that the disk total mass in H$_2$ is $0.025 M_{\odot}$ \citep{Calahan2021}.
Since we only explore a disk face-on, we assumed a gas temperature profile that follows a radial power law with values adapted to the disk upper layers \citep{Huang2018}, without inclusion of a vertical temperature gradient:
\begin{equation}
    T_{\rm gas} = T_{\rm atm} = T_{\rm atm0}\left(\frac{r}{10 \rm \, au} \right)^{-q},
\end{equation}
where we set $T_{\rm atm0}=125$ K and $q=0.47$ following \citealt{Huang2018}. Our calculations do not include dust in the disk. 

We computed $^{12}\rm CO(2-1)$ line radiative transfer predictions, under local thermodynamical equilibrium (LTE) assumption, centered at a frequency of 230.538 GHz. The disk layer probed by $^{12}\rm CO(2-1)$ in our models is at approximately two pressure scale heights from the disk midplane.
For the molecular data, we used that provided by the Leiden LAMDA database \citep{Schroier2010}. The computed synthetic data cubes have a velocity resolution of 40 m s$^{-1}$. To explore the effect of spatial resolution in our predictions, we computed simulated ALMA observations with the \textsc{CASA} software \citep{McMullin2007} version 6.6. Using \textsc{CASA simobserve} we simulate observations that combine configurations C-1, C-4 and C-7, with longest baselines of 161 m, 784 m and 3.6 km, respectively. We used the task \textsc{tclean}, applying \textsc{uvtaper} to produce images with a synthesized beam that has a full width half maximum (FWHM) of $\sim 0.18$ arcsecond. The resulting beam shape is $0.191\times 0.173$ arcsecond with a PA of $-84.9^\circ$.
The spectral and spatial resolutions of our simulated observations are comparable to that achieved in \cite{Teague2022}. We do not consider the effect of thermal noise in the simulated observations.
Finally, we compute the centroid of the Doppler-shifted line emission (line-of-sight velocity) at each image pixel with \textsc{bettermoments} \citep{Teague2018}, by collapsing the cube using the intensity weighted average velocity (first moment).

\begin{figure*}
    \centering
    \includegraphics[width=1\textwidth]{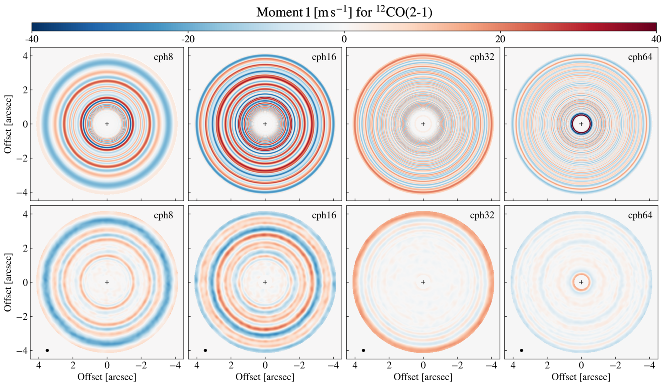}
    \caption{Maps of the intensity weighted average velocity (Moment 1) of $^{12}\rm CO(2-1)$ line emission predictions computed from our set of numerical simulations \texttt{cph}. Top: full resolution image. Bottom: simulated ALMA observation with a synthesized beam of $0.191\times 0.173$ arcsec, shown at the lower left corner of each panel.
    }
    \label{fig:obs}    
\end{figure*}

In Figure \ref{fig:obs}, we show the first moment maps of our raw data cubes (first row), and simulated observations (bottom row). From left to right are predictions computed from fiducial models \texttt{cph}, from lowest to higher grid resolutions up to 64 cells per $H$. Higher than that, the corrugation mode wavelengths converge (Figure \ref{fig:wn}, left panel). It is clear that the observational signatures from VSI are significantly washed out for grid resolution higher than 32 cells per $H$. The dilution of the signal from perturbations is due to the small radial wavelength of VSI corrugation modes relative to the synthesized beam FWHM of the simulated observations. Therefore, it is crucial to understand the convergence of the radial corrugation wavelengths in global high-resolution numerical simulations, as it may have a substantial impact on the observability of the VSI kinematic signatures. 

Molecular line emissions have constrained flared disks with higher values of aspect ratio outside 100 au \citep[see e.g.,][]{zhang2021}. We note that if the disk aspect ratio is on the order of $h\sim0.1$, the above results may underestimate the observability of corrugation modes, as wavelengths in models \texttt{cph\_hr0.1} converge to a value close to that of \texttt{cph16}. Furthermore, if the disk undergoes viscous processes that can coexist with the VSI, the corrugation modes feature may also become detectable, depending on the nature and strength of the viscosity.

\section{Conclusions and Discussion}\label{sec:cd}

In this paper, we study the convergence of radial wavelength of VSI corrugation modes with grid resolution. To this end, we conduct a set of global 2D simulations with Athena++ for different grid resolutions. The resolutions span from 8 cells per $H$ up to 128 cells per $H$. Apart from the fiducial models \texttt{cph} ($\nu=0$ and $h=0.05$), we also carry out simulations with higher aspect ratio ($h=0.1$; \texttt{cph\_hr0.1}) and non-zero viscosity ($\nu=10^{-6}$; \texttt{cph\_visc}), to extend the conclusion such that it applies to flared and viscous disks. We summarize the main findings as follows:

\begin{itemize}

\item Our simulations show that convergence is emerged at 64 cells per $H$ for fiducial models \texttt{cph}. Below it, the radial wavelength decreases with grid resolution.

\item Similar pattern of convergence is observed with higher disk aspect ratio $h=0.1$ (\texttt{cph\_hr0.1}). The saturated radial wavelength is, however, wider than models \texttt{cph} as predicted by linear theory.

\item A small viscosity of $\nu=10^{-6}$ tends to smooth out short radial wavelengths. All \texttt{cph\_visc} models saturate into the same wavelengths for $R>5.5$, and the saturated wavelength is wider than models \texttt{cph}.

\end{itemize}

To inspect how the radial wavelength impacts the observables, we generate the synthesized line observations of $^{12}\rm CO(2-1)$ for a perfectly face-on, TW Hydrae like protoplanetary disk, utilizing fiducial models \texttt{cph} data. The mock observations show that,

\begin{itemize}

\item With resolutions greater than 32 cells per $H$, the signature of corrugation modes is significantly washed out. Therefore, if real disks contain a relative small aspect ratio $h\sim0.05$, we face significant difficulties identifying the VSI from molecular line observations. 

\item Flared disks with $h\sim0.1$ as well as disks undergo viscous processes that can coexist with VSI, can have potential better observability.

\end{itemize}

The best chance to detect VSI kinematic signatures are at the upper layers of the disk, traced by $^{12}$CO, where the VSI velocity perturbations are the strongest, and of the outermost regions of flared protoplanetary disks, where the physical size of corrugation modes feature is largest. In deeper disk layers traced by other CO isotopologues such as $^{13}$CO and C$^{18}$O, the velocity magnitudes of the VSI-induced perturbations are weaker (see e.g. Figure A.4 in \citealt{marcelo+21}). In addition, due to their lower abundance, high-resolution ALMA observations of $^{13}$CO and C$^{18}$O require longer integrations to reach the signal-to-noise needed for kinematic analysis (e.g., \citealt{Teague2021}), while also being affected by projection effects from tracing a larger column of gas (see \citealt{Pinte2023}). 

Alternatively, the meridional flows induced by the VSI can induce an observational signature in the mm-dust continuum emission \citep{flock_etal17}. If the VSI is active near the midplane layers of the disk, its meridional perturbations can drive vertical stirring of fairly coupled dust grains, significantly increasing the disk dust scale height \citep{sk16, flock_etal17, Lin19, flock_etal20, Lehmann2022, Dullemond2022}. The vertical thickness of the disk mm-dust emission, observable with high-resolution continuum ALMA observations (e.g., \citealt{pinte_etal16, Villenave2020, Doi2021, Villenave2022}), can then be used as a diagnostic of the presence of VSI motions \citep{flock_etal17, Dullemond2022}. If the wavelength of the corrugation flows does not significantly alter the effect of VSI dust-gas dynamics in the disk mm-dust scale height relative to previous works, resolving the vertical dust scale height is still an effective approach to detecting VSI signatures in protoplanetary disks. Nevertheless, further high-resolution multi-fluid dust and gas simulations of VSI-unstable disks need to be conducted to confirm our predictions.

From Figure \ref{fig:wn}, it shows that a sautration for corrugation modes is achieved at 64 cells per scale height for fiducial simulations. The wave-wave interactions proposed in \citet{cl22} indicated that if $k_x/k_z=10$, it requests $\sim 20-30$ times higher grid resolution to resolve the parametric instability among the VSI and inertial waves. Such a high resolution is extremely computationally prohibitive and has never been reached in the previous works. Even for higher aspect ratio of $h=0.1$ or non-zero viscosity, where saturation can be achieved with 32 cells per scale height, it still requires massive computational resources to conduct the global simulation of the parametric instability. Therefore, we remain optimistic about the proposed inertial-wave resonance interactions as a potential final state of the VSI, and look forward to future global simulations capable of achieving such high resolutions.

\section*{Acknowledgements}

The authors thank Shangjia Zhang for the helpful discussions. YD acknowledges support from David Brink fund from Balliol College, University of Oxford. CC acknowledges funding from STFC grant ST/T00049X/1 and NSERC. Numerical simulations are conducted on the FAWCETT and CSD3 clusters, University of Cambridge. 

\section*{Data Availability}

The data underlying this article will be shared on reasonable request to the corresponding author.


\appendix


\bibliographystyle{mnras}
\bibliography{vsi} 


\bsp	
\label{lastpage}
\end{document}